\documentstyle[aps,epsfig,prl,multicol]{revtex}
\begin{document}

\draft

\title{Selective Transparence of Single-Mode Waveguides with Surface
Scattering}

\author{F.~M.~Izrailev$^1$ and N.~M.~Makarov$^2$}

\address{$^1$Instituto de F\'{\i}sica, Universidad Aut\'{o}noma de Puebla,
        Apartado Postal J-48, Puebla, Pue., 72570, M\'{e}xico}

\address{$^2$C.I.D.S., Instituto de Ciencias, Universidad Aut\'{o}noma
         de Puebla, \\ Priv. 17 Norte No 3417, Col. San Miguel
         Hueyotlipan, Puebla, Pue., 72050, M\'{e}xico }

\date{\today}

\maketitle

\begin{abstract}
A random surface scattering in a one-mode waveguide is studied in
the case when the surface profile has long-range correlations
along the waveguide. Analytical treatment of this problem shows
that with a proper choice of the surface, one can arrange any
desired combination of transparent and non-transparent frequency
windows. We suggest a method to find such profiles, and
demonstrate its effectiveness by making use of direct numerical
simulations.
\end{abstract}

\begin{multicols}{2}

Recently much attention is paid to 1-dimensional solid state
models with a correlated disorder. The interest to this subject is
due to the results that demonstrate a direct relevance of
anomalous transport properties to particular pair correlations in
the cite potential. Specifically, it was shown
\cite{IK99,IKU01} that any wanted combination of
transparent and non-transparent frequency intervals can be
realized by a proper construction of potentials with long-range
correlations. Experimental realization of such potentials for
single-mode waveguides with inserted delta-like scatters
\cite{KIKS00} has proved a possibility to
construct devices that are transparent for any given range of
frequency.

On the other hand, there is a well developed theory of a wave
propagation through surface-disordered waveguides (see, e.g.,
\cite{appl,SFMY9899} and references therein). This subject is important
both from the theoretical viewpoint, and for experimental
applications, such as the optical fibers, remote sensing, radio
wave propagation, shallow water waves etc., \cite{appl}. One
should note, that main results in this field refer to random
surfaces with a {\it fast} decay of correlations along a scattering
surface. It is now of great interest to explore the role of {\it slowly}
decaying correlations that are expected to result in anomalous
properties of surface scattering.

We would like to stress that the problem of a correspondence
between surface and bulk scattering still deserves a detailed
study. In this respect, one should mention recent results
\cite{SFMY9899} where specific properties of a surface scattering have
been discovered, that are different from those known in standard
quasi-1D models with random potentials \cite{FM94}.

In this Letter we analyze a surface scattering in the case of one
open channel, assuming long-range correlations in the surface
potential. Our main interest is to explore a possibility to
construct such surfaces that result in windows of transparence in
dependence on the frequency of incoming waves.

We consider a 2D waveguide of the length $L$ along the
$x$-axis, with a flat upper surface $z=d$ and a rough (corrugated)
lower surface $z=\xi(x)$, see also
\cite{SFMY9899,MakYur89FrMakYur90}. The random function $\xi(x)$ is
characterized as follows,
\begin{equation}
\langle\xi(x)\rangle=0, \qquad
\langle\xi(x)\xi(x')\rangle=\sigma^2{\cal W}(|x-x'|).
\label{ksi}
\end{equation}
The angular brackets stand for a statistical average over the
ensemble of realizations of $\xi(x)$ and the root-mean-square
$\sigma$ determines the roughness strength. The binary correlator
${\cal W}(|x|)$ decreases with a typical scale $R_c$ which is of
the order of a mean length of surface irregularities. Obviously, a
statistical treatment of the surface scattering is meaningful only
if the correlation length $R_c$ is much less than the length size
$L$ of a waveguide, $R_c\ll L$. In addition, we assume that the
roughness is weak, $\sigma\ll d$.

Keeping in mind the relevance of a surface scattering in our model
to the Anderson localization, in what follows we consider a
single-mode waveguide when there is the only propagating mode with a
real longitudinal wave number $k_1=\sqrt{(\omega/c)^2-(\pi
/d)^2}$. All other modes are evanescent, therefore, the width $d$
is restricted by the conditions, $0<k_1d/\pi<\sqrt{3}$. Note that
the assumed condition $\sigma\ll d$ leads to the inequality
$k_1\sigma\ll 1$.

It can be shown~\cite{MakYur89FrMakYur90} that the corresponding
wave equation takes the form,
\begin{equation}\label{Fel-eq}
\left[\frac{d^2}{dt^2}+\left(k_1R_c\right)^2 \right]
\Psi(t)=\frac{2}{\pi}\,\frac{\sigma}{R_c}\,
\left(\frac{\pi R_c}{d}\right)^3\,\varphi(t)\,\Psi(t)
\end{equation}
with $t=x/R_c$. Here the function $\varphi(t)$ is determined by
the relation $\xi(x)=\sigma\,\varphi(x/R_c)$. One can see that the
problem of surface scattering in a single-mode waveguide is
reduced to a 1D-model with random potential
$\varphi(t)$. Therefore, its solution is entirely consistent with
the theory of 1D localization. In accordance with the form of the
perturbation potential, the backscattering length $L_b$ is given by
the following expression,
\begin{equation}\label{Lb}
L_b^{-1}=\frac{4\sigma^2}{\pi^2}\left(\frac{\pi}{d}\right)^6
\frac{W(2k_1)}{(2k_1)^2}.
\end{equation}
Here $W(k_x)$ is the Fourier transform of a binary correlator
${\cal W}(|x|)$, ${\cal
W}(|x|)=\int_{-\infty}^{\infty}\frac{dk_x}{2\pi}
\exp\left(ik_xx\right)\,W(k_x)$. Note that $W(k_x)$ is a positive
function of the order of $R_c$, which decreases in dependence on
$|k_x|$ with a typical scale $R_c^{-1}$.

The expression (\ref{Lb}) for the backscattering length $L_b$
gives a complete information about the transmission through the
waveguide. In particular, it determines the conductance and its
fluctuations for any ratio $L/L_b$, see e.g. \cite{LGP88}. As one
can see, the binary correlator of the surface profile defines
all transport properties of a one-mode waveguide. In particular,
if $W(2k_1)$ vanishes within some interval of the wave number
$k_1$, the waveguide is fully transparent. Below, we show how to
construct surface profiles that result in a complete
transparence in a given range of $k_1$.

To do this, we generalize the approach \cite{IKU01} developed for
the Kronig-Penney model with the correlated disorder. This model
can be treated as a particular case of the equation (\ref{Fel-eq})
when the function $\varphi(t)$ is given in the form of periodic
delta-kicks with random amplitudes.

In the case of continuous potential the problem can be solved by
representing  $\varphi(t)$ in the following form,
\begin{equation}
\varphi(t)=\int_{-\infty}^\infty\,dt'\, Z(t-t')\,\beta(t');
\label{phi-def}
\end{equation}
with some function $\beta(t)$, which is entirely determined by the
Fourier transform $W(k_x)$ of the binary correlator ${\cal
W}(|x|)$,
\begin{equation}
\beta(t)=\sqrt{R_c}\,\int_{-\infty}^{\infty}\frac{dk_x}{2\pi}
\exp\left(ik_xR_ct\right)\,W^{1/2}(k_x).
\label{beta-def}
\end{equation}
Here $Z(t)$ is a dimensionless delta-correlated random process
(white noise) with $\langle Z(t)\rangle=0$ and $\langle
Z(t)Z(t_0)\rangle=\delta(t-t_0)$.

The above expressions allows one to practically solve an inverse
scattering problem of constructing a potential from its binary
correlator. Note, that this construction is possible in the case
of a weak disorder, $\sigma \ll d$, this is why only binary
correlator is involved in the reconstruction of $\varphi(t)$. For
this reason, the above solution is not unique since higher
correlators are not controlled. Below we demonstrate the suggested
approach by considering two simple cases of a correlated surface
profile.

Let us first consider the case when the waveguide is completely
transparent for $k_1 > 1/2R_c$. In this case one can get the
following expressions for the binary correlator and its Fourier
transform,
\begin{eqnarray}
{\cal W}_1(|x|)&=&\frac{\sin(x/R_c)}{x/R_c};
\label{W1-exp} \\
W_1(k_x)&=&\pi R_c\,\Theta(1-|k_x|R_c)
\label{FT-W1}
\end{eqnarray}
with $\Theta(x)$ as the unit-step function, $\Theta(0)=1/2$. 
This kind of a $\Theta$-like dependence of $W(k_x)$ was recently analyzed 
in \cite{SLM00}. The surface profile that has the above correlations, 
is described by the function,
\begin{equation}
\varphi_1(t)=\frac{1}{\sqrt{\pi}}\,\int_{-\infty}^\infty\,
dt'\,Z(t-t')\,\frac{\sin t'}{t'}.
\label{phi1}
\end{equation}
In this case the inverse backscattering length is
\begin{equation}
\label{L1}
\frac{1}{L_{b1}(k_1)}=\frac{4}{\pi R_c}
\left(\frac{\sigma}{R_c}\right)^2
\left(\frac{\pi R_c}{d}\right)^6
\frac{\Theta(1-2k_1R_c)}{\left(2k_1R_c\right)^2}.
\end{equation}
Therefore, with an increase of the wave number $k_1$ the
backscattering length smoothly increases, and after, goes abruptly
to infinity for $k_1>1/2R_c$. Such a behavior can be observed if
the transition point $k_1=1/2R_c$ is located inside an allowed
single-mode region, for  $12\,\left(\pi R_c/d\right)^2>1$, see
above. In order to observe this effect for finite waveguides, one
needs to assume that at the transition point the regime of a
strong localization holds,
\begin{equation}\label{loc1}
\frac{L}{L_{b1}(1/2R_c)}\equiv\frac{2}{\pi}\,
\left(\frac{\sigma}{R_c}\right)^2\,
\left(\frac{\pi R_c}{d}\right)^6\,
\frac{L}{R_c}\gg 1.
\end{equation}
In this case the average transmittance is expected to be
exponentially small due to a strong localization for $k_1<1/2R_c$.
In contrast, in the interval
$1<\left(2k_1R_c\right)^2<12\,\left(\pi R_c/d\right)^2$ a
ballistic regime occurs with a perfect transparence.

The second case refers to a complimentary situation when for
$k_1<1/2R_c$ the waveguide is transparent and for $k_1>1/2R_c$ is
non-transparent. One can find that the corresponding expressions
for ${\cal W}(|x|)$ and $W(k_x)$ are
\begin{eqnarray}
{\cal W}_2(|x|)&=&\pi\delta(x/R_c)-\frac{\sin(x/R_c)}{x/R_c};
\label{W2-exp} \\
W_2(k_x)&=&\pi R_c\,\Theta(|k_x|R_c-1).
\label{FT-W2}
\end{eqnarray}
In this case the corrugated surface is described by a
superposition of the white noise and the roughness of the first
type. As a result, the surface-profile potential $\varphi(t)$
takes the form,
\begin{equation}
\varphi_2(t)=\sqrt{\pi}\,Z(t)-
\frac{1}{\sqrt{\pi}}\,\int_{-\infty}^\infty\,
dt'\,Z(t-t')\,\frac{\sin t'}{t'}.
\label{xi2}
\end{equation}
Correspondingly, the inverse backscattering length is expressed by
\begin{equation}\label{L2}
\frac{1}{L_{b2}(k_1)}=\frac{4}{\pi R_c}
\left(\frac{\sigma}{R_c}\right)^2
\left(\frac{\pi R_c}{d}\right)^6
\frac{\Theta(2k_1R_c-1)}{\left(2k_1R_c\right)^2}.
\end{equation}

Therefore, in contrast to the first case, the backscattering
length $L_{b2}(k_1)$ is equal to infinity below the transition
point $k_1=1/2R_c$. With further increase of $k_1$, it smoothly
increases starting from the value $L_{b2}(1/2R_c)$. Again, the
transition point $k_1=1/2R_c$ is supposed to be inside the
single-mode interval. In addition, we assume a strong localization
to be retained at the upper point $k_1=\pi\sqrt{3}/d$ of the
single-mode region, i.e.,
\begin{equation}\label{loc2}
\frac{L}{L_{b2}(\pi\sqrt{3}/d)}\equiv\frac{1}{3\pi}\,
\left(\frac{\sigma}{R_c}\right)^2\,
\left(\frac{\pi R_c}{d}\right)^4\,\frac{L}{R_c}\gg 1.
\end{equation}
In this case the ballistic transparence is abruptly replaced by a
strong localization at the transition point $k_1=1/2R_c$.

Let us now demonstrate the above predictions by direct numerical
simulations. Keeping in mind the relation $4L_b=l_\infty$ between
the backscattering length $L_b$ and the localization length
$l_\infty$ in infinite waveguides (for $L \rightarrow \infty$), we
compute $l_\infty$ by the method described in \cite{IK99}. For
this, we approximate the continuous function $\varphi(t)$ in
Eq.(\ref{Fel-eq}) by the sum of delta-kicks with the spacing
$\delta$ chosen much smaller that any physical length scale in our
model. In this way one can write a Hamiltonian map which can be
used to find the localization length via the Lyapunov exponent
$\lambda=l_\infty^{-1}$ of a dynamical problem associated with
this map (see details in \cite{IK99,IKU01}).

\begin{figure}[htb]
\vspace{-1.cm}
\begin{center}
\hspace{-2.3cm}
\epsfig{file=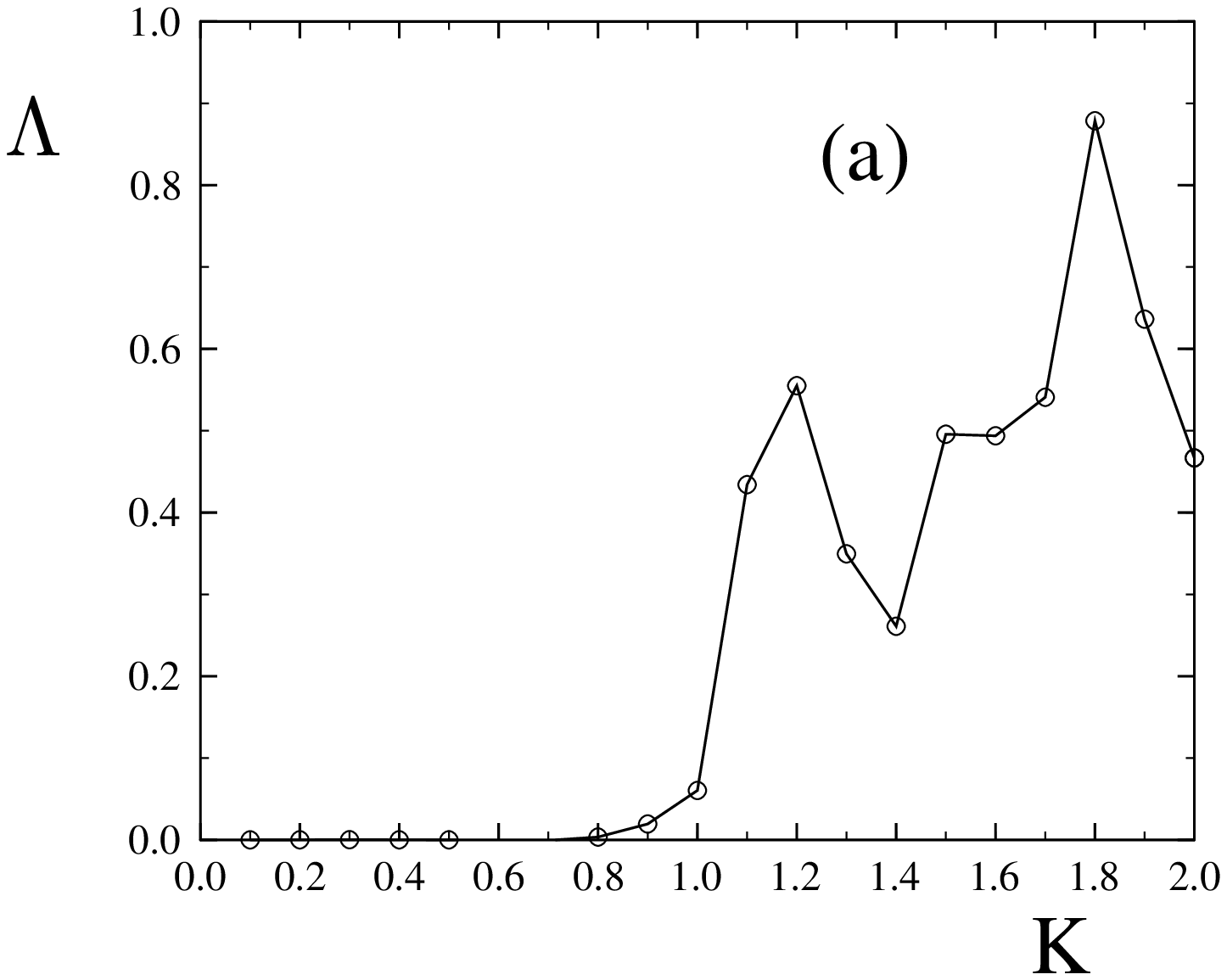,width=2.2in,height=2.4in,angle=-90}
\vspace{-0.5cm}
\end{center}
\end{figure}

\begin{figure}[htb]
\vspace{-3.cm}
\begin{center}
\hspace{-2.3cm}
\epsfig{file=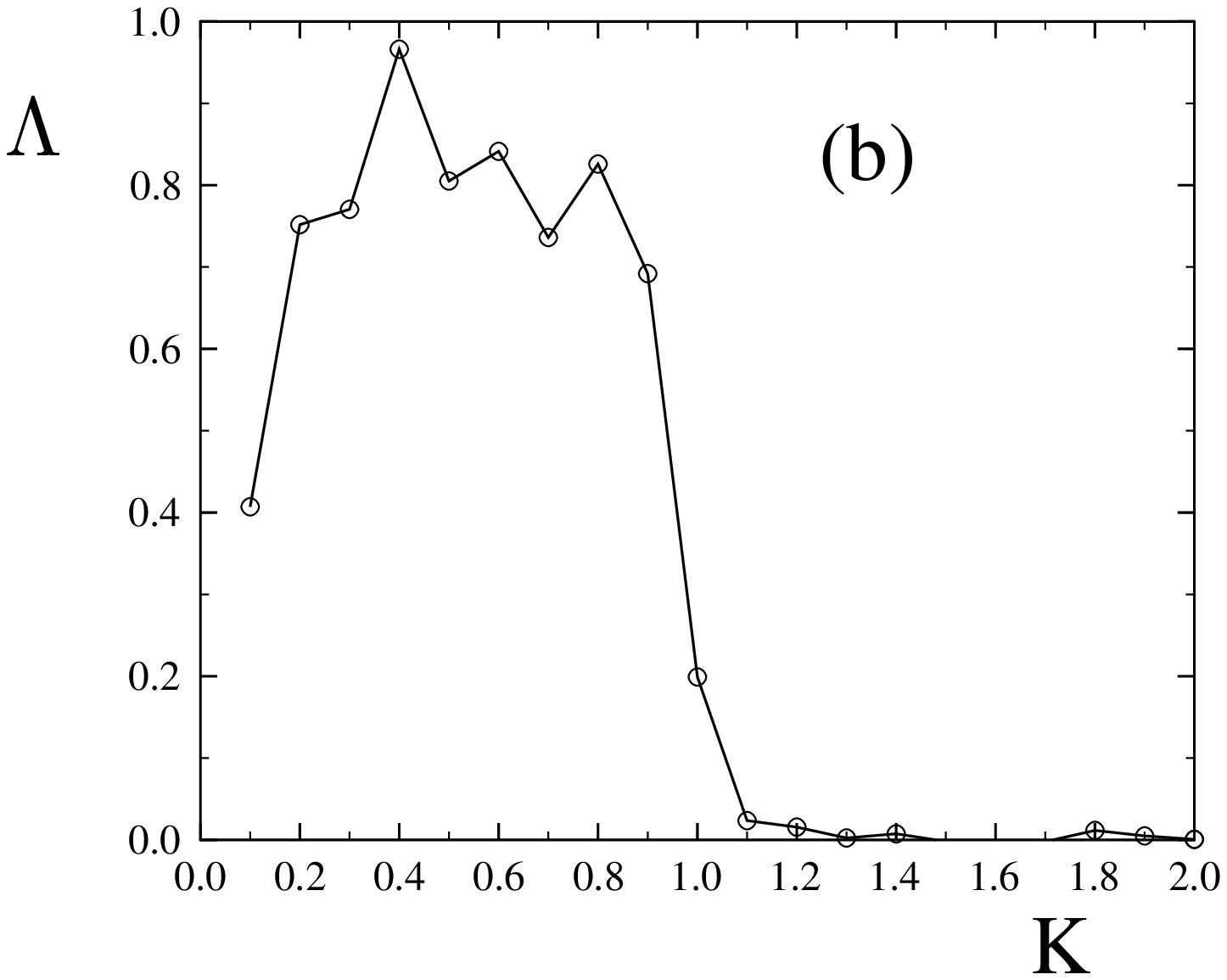,width=2.2in,height=2.4in,angle=-90}
\vspace{-0.5cm}
\narrowtext
\caption{Selective dependence of the rescaled localization length
on the normalized wave vector for two realizations of a random 
surface with long-range correlations, see in the text. In both 
cases the value of $c_0$ was chosen in order to have the same 
scale for $\Lambda$.}
\end{center}
\end{figure}

Numerical data are reported in Fig.1 where the normalized Lyapunov
exponent $\Lambda=c_0\lambda K^2$ is plotted against the
normalized wave vector $K=2k_1R_c$ for the range $0<K<2$ which
corresponds to the single-mode interval. In these units one can
clearly see a non-trivial dependence of $\lambda$ on the wave
vector, which is due to specific binary correlations in the
potential $\varphi(t)$.

The potential $\varphi(t)$ was constructed according to discrete
versions of the expressions (\ref{phi1}) and (\ref{xi2}) which
determine complimentary dependences of the backscattering length
$L_b(k_1)$, see (\ref{L1}) and (\ref{L2}). The data clearly
demonstrate a drastic dependence of $\Lambda$ on the wave vector
when crossing the transition point $K=1$. By taking the size $L$
in accordance with the expressions (\ref{loc1}) and (\ref{loc2}),
one can arrange the selective transparence of waveguides as is
predicted by the theory.

In conclusion, we study the possibility to construct one-mode
waveguides with a selective transparence in dependence on the wave
vector of an incoming wave. Analytical treatment shows that this
can be done by a proper choice of random surfaces with
specific long-range correlations along waveguides. Numerical
data for two cases with complimentary dependences of the
backscattering length on the wave vector demonstrate the
effectiveness of the theoretical predictions. The results
presented here may be used for experimental realizations of
waveguides with a desired selectivity of the transmission. 
Note that random surfaces with discontinuous dependence of $W(k_x)$ 
have been recently fabricated in the experimental 
study of backscattering enchancement \cite{WOD95}. 

This work was supported by the CONACyT (Mexico) Grant No. 34668-E.
N.M.M. acknowledges support from CONACYT.

\end{multicols}

\end{document}